\documentclass{article} 
\usepackage{arxiv}

\usepackage{hyperref}
\usepackage{url}

\usepackage{graphicx} 

\usepackage{mathtools}
\usepackage{amssymb}

\usepackage{algorithm}
\usepackage{algpseudocode}
\usepackage{wrapfig}
\usepackage{subcaption}

\usepackage{natbib}
\title{Reinforcement Networks: \\
novel framework for collaborative \\
Multi-Agent Reinforcement Learning tasks}

\author{
Maksim Kryzhanovskiy$^{1,2}$,
Svetlana Glazyrina$^{2}$,
Roman Ischenko$^{1,2}$,
Konstantin Vorontsov$^{1,2}$
\\[0.5em]
$^{1}$Institute for Artificial Intelligence, Lomonosov Moscow State University \\
$^{2}$Lomonosov Moscow State University
}

\begin{document}

\maketitle

\begin{abstract}

Modern AI systems often comprise multiple learnable components that can be naturally organized as graphs. A central challenge is the end-to-end training of such systems without restrictive architectural or training assumptions. Such tasks fit the theory and approaches of the collaborative Multi-Agent Reinforcement Learning (MARL) field. We introduce Reinforcement Networks, a general framework for MARL that organizes agents as vertices in a directed acyclic graph (DAG). This structure extends hierarchical RL to arbitrary DAGs, enabling flexible credit assignment and scalable coordination while avoiding strict topologies, fully centralized training, and other limitations of current approaches. We formalize training and inference methods for the Reinforcement Networks framework and connect it to the LevelEnv concept to support reproducible construction, training, and evaluation. We demonstrate the effectiveness of our approach on several collaborative MARL setups by developing several Reinforcement Networks models that achieve improved performance over standard MARL baselines. Beyond empirical gains, Reinforcement Networks unify hierarchical, modular, and graph-structured views of MARL, opening a principled path toward designing and training complex multi-agent systems. We conclude with theoretical and practical directions — richer graph morphologies, compositional curricula, and graph-aware exploration. That positions Reinforcement Networks as a foundation for a new line of research in scalable, structured MARL.
    
\end{abstract}

\section{Introduction}

Modern AI systems increasingly comprise multiple learnable components that must coordinate to solve complex tasks. Prominent examples include large-language-model (LLM) workflows for tool use and orchestration, Retrieval-Augmented Generation (RAG) pipelines, and multi-agent LLM systems. These systems are naturally expressed as \emph{graphs} of interacting modules: nodes represent learnable components (e.g., retrievers, planners, controllers), and edges encode information flow and control dependencies. This view is well-grounded in the literature on graph-structured learning and computation, where graphs serve as a unifying abstraction for modular architectures and message passing \cite{zhou2020graph,khemani2024review}. In practice, real-world workflows are frequently modeled as \emph{directed acyclic graphs} (DAGs), which make dependencies explicit, enable topological scheduling, and support scalable orchestration in scientific and data-intensive computing \cite{verucchi2023survey}. Recent LLM systems adopt precisely these ideas: LLM-driven workflow generators and orchestrators (e.g., WorkflowLLM; automated DAG construction for enterprise workflows) formalize complex LLM pipelines as DAGs for reliable execution and optimization \cite{fan2025workflowllm, xu2024llm4workflow}, and DAG-structured plans have been shown to improve task decomposition and concurrency in embodied LLM-agent settings \cite{gao2024dag}. RAG, now a standard pattern for knowledge-intensive tasks, is also naturally modular — combining retrievers, rerankers, and generators in a graph that can be adapted and optimized end-to-end \cite{gao2023retrieval,gupta2024comprehensive,zhao2024retrieval}.

\paragraph{From learnable systems to collaborative MARL.}
Graph-structured AI pipelines consist of interacting modules whose objectives and behaviors naturally motivate viewing them as \emph{agents} with (partially) aligned goals under partial information and non-stationarity—the core setting of cooperative multi-agent reinforcement learning (MARL). Foundational work in cooperative MARL formalizes centralized training for decentralized execution (CTDE), credit assignment, and coordination as central challenges \cite{amato2024introductioncentralizedtrainingdecentralized, huh2024multiagentreinforcementlearningcomprehensive}. Credit-assignment methods such as LICA show that joint optimization is possible without fully centralized execution \cite{zhou2020learningimplicitcreditassignment}, while hierarchical reinforcement learning (HRL) offers temporal abstraction and multi-level decision-making \cite{sutton1999between}. Together, these insights suggest a unified view: graph-structured, learnable systems as collaborative MARL problems where nodes (modules/agents) coordinate over a computation graph.
\paragraph{Leveled architectures and the LevelEnv abstraction.}
The \emph{LevelEnv} abstraction from the TAG framework \cite{paolo2025tag} operationalizes this view for decentralized hierarchical MARL by treating each hierarchy level as the “environment’’ for the level above, standardizing information exchange while preserving loose coupling. This supports deep, heterogeneous hierarchies, improves sample efficiency and performance on standard MARL benchmarks, and avoids rigid two-level manager–worker designs, aligning closely with modern LLM workflows and RAG pipelines \cite{paolo2025tag}.
\paragraph{Why DAGs for multi-agent systems?}
Modeling multi-agent, multi-module systems as DAGs enables topological ordering and parallelization of independent subgraphs \cite{verucchi2023survey}, clarifies credit-assignment paths along directed edges and aligns with hierarchical task/dependency and HTN formulations \cite{georgievski2015htn, chen2021learning}, and avoids deadlocks and circular dependencies in training and execution \cite{verucchi2023survey}. DAGs also generalize trees, supporting shared substructures and multi-parent dependencies common in tool-using LLM agents and modular RAG systems \cite{fan2025workflowllm, xu2024llm4workflow, gao2023retrieval}.

\paragraph{Our research.}
We build on these observations to introduce \emph{Reinforcement Networks}, a general framework that organizes collaborating agents as nodes in a DAG. Our formulation unifies hierarchical, modular, and graph-structured views of MARL: it supports flexible credit assignment and scalable coordination without strict topologies or fully centralized training. Moreover, it connects directly to the LevelEnv interface \cite{paolo2025tag} for reproducible construction, training, and evaluation. We show that DAGs outperform the tree structures used in TAG for collaborative tasks. We highlight future directions—richer graph structures, compositional curricula, and graph-aware exploration—for scalable, structured MARL.

\section{Related Works}

\paragraph{Multi-Agent Reinforcement Learning}
Multi-agent systems have seen rapid growth, driven by autocurricula emerging from interacting learning agents and enabling continual improvement \citep{nguyen2020deeprlma, oroojlooy2023coopmadrl}. Tooling such as PettingZoo \citep{terry2021pettingzoo} and BenchMARL \citep{bettini2024benchmarl} standardizes environments and benchmarks, improving comparability and reproducibility.
Independent learning methods treat each agent as solving a partially observable RL problem where others are part of the environment. Examples include IPPO \citep{dewitt2020ippo}, IQL \citep{tan1997marl}, and ISAC \citep{bettini2024benchmarl}, extending PPO \citep{schulman2017ppo}, Q-Learning \citep{watkins1992ql}, and SAC \citep{haarnoja2018sac}.
Parameter-sharing architectures rely on shared critics or value functions, as in MAPPO \citep{yu2022mappo}, MASAC \citep{bettini2024benchmarl}, and MADDPG \citep{lowe2017maddpg}.
Explicit communication methods enable inter-agent information exchange via consensus schemes \citep{cassano2020mavaluefun} or learned communication protocols \citep{foerster2016communic, jorge2016communic}, directly addressing coordination.
A core challenge in MARL is non-stationarity from simultaneous policy updates, which quickly invalidates replay data \citep{foerster2016communic}. Centralized Training with Decentralized Execution (CTDE) partly mitigates this via shared training components \citep{oroojlooy2023coopmadrl}, but its constraints limit applicability in lifelong learning settings requiring continuous adaptation.
\paragraph{Hierarchical Reinforcement Learning}
Hierarchy enables abstraction-based value propagation and temporally and spatially extended behavior, improving exploration and efficiency over flat RL \citep{hutsebaut2022hrlchal, nachum2019hrladv}. Decomposing tasks also reduces computational complexity and promotes sub-problem reuse, accelerating learning.
Various approaches instantiate HRL. The Options framework models temporally extended actions via SMDPs (“options’’) with policies, termination conditions, and initiation sets \citep{sutton1999between}, later trained end-to-end in Option-Critic \citep{bacon2017optcri}. Feudal RL instead uses a manager–worker hierarchy, where high-level managers issue intrinsic goals to lower-level workers \citep{dayan1992frl, vezhnevets2017frl}. A core challenge is that changing low-level policies induces non-stationarity in higher-level value estimation, motivating model-based methods such as CSRL \citep{li2017modelbasedhrl}.
Combining HRL with MARL adds further non-stationarity through multi-agent dependencies. The TAG framework \citep{paolo2025tag} proposes methods to mitigate these issues but remains closely tied to the TAME formulation \citep{levin2021tame}, constraining architectures to layered digraphs. Our work extends this line by relaxing these structural limitations.

\section{Methodology}

\subsection{Fundamentals}

\paragraph{Multi-Agent Reinforcement Learning}
A MARL setting is defined by an environment with state space $S^{env}$, agents ${w_i}{i=1}^k$, a joint action space $A$ with individual spaces $A_i$, a transition function $p: S^{env} \times A \rightarrow \Delta(S^{env})$, an initial-state distribution $p_0 \in \Delta(S^{env})$, and rewards ${R_i}{i=1}^{k}$, where $r_i^{env}: S^{env} \times A_i \rightarrow \mathbb{R}$ is the reward of the $i$-th agent.

\subsection{Proposed model}

\paragraph{Reinforcement Networks}

\begin{wrapfigure}{r}{0.4\textwidth}
    \begin{center}
    \includegraphics[width=0.3\textwidth]{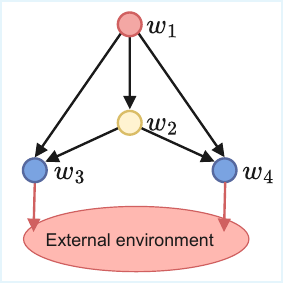}
    \end{center}
    \caption{Considering agent $\omega_2$ colored yellow, its subordinate agents $V_2^-$ are marked blue and the superior agents are marked red. Edges of the DAG represent agents' couples where direct communication is present. Red arrows depict interaction between \textit{motors} and the real environment.}
    \label{fig:dag_example}
\end{wrapfigure}

Consider a directed acyclic graph $G = (V, E),$ where $V = \{w_1, \ldots, w_N\}$ stands for the set of vertices representing the agents, and $E$ is the set of directed edges. We introduce the following notation to formally characterize the local connectivity of each vertex $w_i \in V$: $I_i^+ = [ j \mid (w_j, w_i) \in E ], \quad I_i^- = [ j \mid (w_i, w_j) \in E ]$. Here, $I_i^+$ denotes the indices list of vertices that have an outgoing edge incident to $w_i$, whereas $I_i^-$ denotes the indices list of vertices that have an incoming edge incident to $\omega_i$. The corresponding cardinalities are designated as $l_i^+ = |I_i^+|, \quad l_i^- = |I_i^-|$.
Furthermore, we define the sets of agents associated with these indices as $V_i^+ = \{\omega_j \mid j \in I_i^+\}, \quad V_i^- = \{\omega_j \mid j \in I_i^-\}$. 

The sink nodes of the graph, for which $V_i^- = \varnothing$, interact directly with the external environment and consequently receive rewards and observations from it. We denote the set of these agents by $V_0 \subseteq V$ and refer to them as \textit{motors}.

In this formulation, we refer to a vertex $w_j$ as a \emph{superior agent} of $w_i$ if there exists an edge $(w_j, w_i) \in E$, and as a \emph{subordinate agent} if there exists an edge $(w_i, w_j) \in E$. This terminology enables a natural hierarchical interpretation of the directed acyclic structure.

For the traversal along an edge, we use the verb \textit{pass}, whereas for the reverse direction we use the verb \textit{return}. Similar to TAG \citep{paolo2025tag}, for a given agent $w_i$ the set of subordinate agents $V_i^-$ plays the role of the environment, returning both the reward and the observation in response to the agent's actions. 

Each agent $w_i$ is represented by a tuple $
\langle M_i, O_i^-, A_i, O_i^+, \pi_i, \phi_i, \psi_i, R_i \rangle$,
which is defined as follows:

\begin{itemize}
    \item $M_i$ - the message space, i.e., the set of messages that the agent may return to its superior agents.
    
    \item $O_i^-$ - the observation space, defined as $O_i^- = \prod_{j \in I_i^-} M_j$, corresponding to the list of messages passed by subordinate agents.
    
    \item $A_i$ - the action space used to influence subordinate agents.
    
    \item $O_i^+$ - the space of instructions, i.e., the actions passed by superior agents, defined as $O_i^+ = \prod_{j \in I_i^+} A_j$.
    
    \item $\pi_i: O_i^- \times O_i^+ \to \Delta(A_i)$ - the agent’s policy. 
    The conditional distribution $\pi_i(a_i \mid o_i^-, a_i^+)$ depends both on the observations obtained from subordinate agents and the instructions provided by superior agents.  
    Introducing the notation $O_i \coloneq O_i^- \times O_i^+$, the policy can equivalently be written as $\pi_i(a_i \mid o_i), \quad o_i \in O_i$.
    
    \item $\phi_i: O_i^- \times \mathbb{R}^{l_i^-} \to M_i$ - the communication function. It determines the message describing current environment state to be transmitted upward. 
    Here, $\mathbb{R}^{l_i^-}$ represents the vector of rewards returned by subordinate agents.  
    Thus, $m_i = \phi_i(o_i^-, r_i^-)$.
    For source vertices with $V_i^+ = \varnothing$, the choice of $\phi_i$ has no impact on the overall system dynamics.
    
    \item $\psi_i: O_i^- \times \mathbb{R}^{l_i^-} \to \mathbb{R}$ - the proxy-reward function. The value of this function is returned to superior agents as an element of the reward list. For source vertices with $V_i^+ = \varnothing$, the choice of $\psi_i$ has no impact on the overall system dynamics.
    
    \item $R_i: \mathbb{R}^{l_i^-} \to \mathbb{R}$ is the aggregation function, which interprets the list of received rewards.
\end{itemize}

We use the following notation throughout:
$$
a_i^+ = [a_j \mid j \in I_i^+], \quad
o_i^- = [m_j \mid j \in I_i^-], \quad
r_i^- = [r_j \mid j \in I_i^-],
$$
where the elements are assumed to be ordered according to related lists $I_i^+, I_i^-$. Finally, $\Delta(A_i)$ denotes the probability simplex over the action space $A_i$.

\subsection{Upstream and downstream inference}

\begin{figure}[h]
    \centering
    \includegraphics[width=0.75\textwidth]{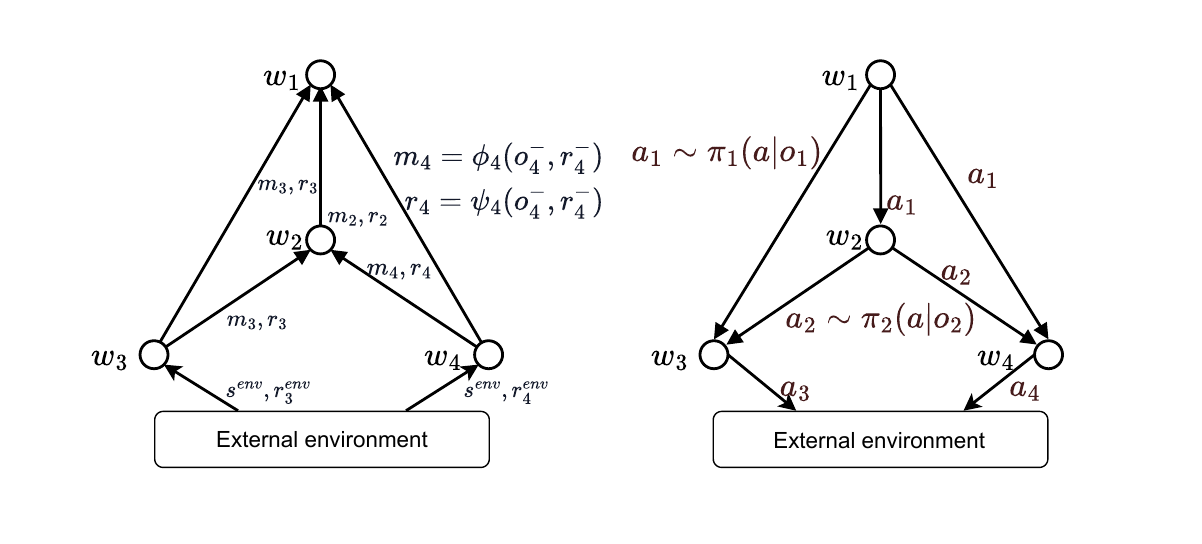} 
    \caption{Example of inference in the system. Left: downstream inference. Right: upstream inference.}
    \label{fig:two_panels}
\end{figure}

\begin{figure}[h]
    \centering
    \includegraphics[width=0.48\textwidth]{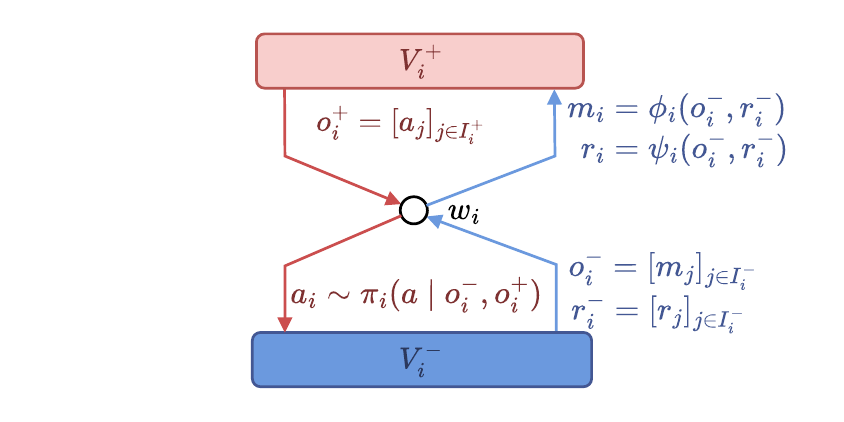} 
    \caption{Information flow from the perspective of agent $w_i$. Rounded rectangles indicate $V_i^+$ (red) and $V_i^-$ (blue). Actions propagate downstream (red), while messages and rewards flow upstream (blue). One timestep corresponds to a counterclockwise cycle starting at the upper-right corner.}
    \label{fig:communication}
\end{figure}

We describe the exchange of information both among agents and between the environment and the agent system. Figure~\ref{fig:two_panels} illustrates the information flow across the entire system, while Figure~\ref{fig:communication} provides a visualization from the perspective of an individual agent. This mechanism generalizes the standard agent-environment interaction in reinforcement learning to hierarchical multi-agent systems. A single timestep of an agent corresponds to one cycle of interaction, beginning with processing observations, followed by sampling an action, and concluding with receiving a new observation as the environment’s response to the chosen action. At each timestep, information flow is decomposed into two phases: upstream inference and downstream inference. Information from the environment — namely observations and rewards — is transmitted to an agent against the direction of outgoing edges, whereas the agent’s action is propagated downwards along the edges of the graph. It is important to note that subordinate agents function as an environment for a given agent. For \textit{motors}, as long as no subordinate agents are present, the external environment assumes this role.

\paragraph{Sharing experience.}

During upstream inference, agents’ observations and rewards are propagated upward through the graph. The functions $\phi$ and $\psi$ generate the messages and rewards that are passed to superior agents. This design enables efficient knowledge sharing. At the same time, it enables partial hiding of the environment state depending on the recipient’s abstraction level or task perspective.

In the beginning of a timestep, agent $w_i$ possesses the information received at the previous step as a response from the environment:
$$
o_i^-(t) = (m_j(t))_{j \in I_i^-},
$$ $$
r_i^-(t - 1) = (r_j(t - 1))_{j \in I_i^-}.
$$

Given that agent $w_i$ generates a message and a reward to be passed to its superior agents:
$$
m_i(t) = \phi_i\big(o_i^-(t), r_i^-(t - 1)\big), 
\qquad 
r_i(t - 1) = \psi_i\big(o_i^-(t), r_i^-(t - 1)\big).
$$
These values then become elements of the input received by the agents of $V_i^+$.
This process propagates experience upward in the hierarchy, enabling higher-level agents to incorporate information from subordinate agents.

\paragraph{Interaction with the environment.}
For a given vertex $w_i$, the set $V_i^-$ serves as the environment for this agent. For \textit{motors} external environment instead of empty sets of subordinate agents is used.

At the start of the second phase of step $t$, which is considered in this paragraph, agent $w_i$ possesses the information received at the end of the previous step from the environment and during the previous phase from its superior agents, respectively:
$$
o_i^-(t) = (m_j(t))_{j \in I_i^-},
$$ $$
o_i^+(t) = (a_j(t))_{j \in I_i^+},
$$
and selects an action according to
$$
a_i(t) \sim \pi_i(a \mid o_i^-(t), o_i^+(t)) = \pi_i(a|o_i(t)).
$$

In response to this action, the environment represented by the set $V_i^-$ returns two vectors: new observations and rewards,
$$
o_i^-(t + 1) = (m_j(t + 1))_{j \in I_i^-}, 
\qquad 
r_i^-(t) = (r_j(t))_{j \in I_i^-}.
$$
All subordinate agents must perform at least one step of execution to generate $m_j(t + 1)$ and $r_j(t)$ for all $j \in I_i^-$.  

To interpret the resulting list of rewards, we utilize the agent's  reward aggregation function $R_i: \mathbb{R}^{l_i^-} \to \mathbb{R}$.
The reward of agent $w_i$ for executing action $a_i(t)$ is then defined as
$$
r_i^{\text{target}}(t) \coloneq R_i(r_i^-(t)).
$$
The function $R_i$ may be specified in various ways (e.g., mean, maximum, weighted sum), providing flexibility and opening avenues for further investigation.

\paragraph{System initialization}

The system inference begins by sampling an initial state $s_0 \sim p_0$ from the external environment. This state is provided as $o_i^-(0)$ to all motors $w_i \in V_0$. We initialize $r_i(-1) = 0$ for all motors. With these initial values, the motors possess all the information required to perform upstream inference at $t = 0$. Once the motors execute, they trigger a cascade of observations and rewards through the superior agents, which in turn return their directives. Using these signals, the motors interact with the external environment and advance to the next timestep.

\paragraph{Multiple time scales.}
It is worth noting that hierarchical graph structure naturally allows different agents to operate on distinct though consistent time scales, reflecting the varying reaction speeds of agents at different depths.  
Specifically, for a single time step of agent $w_i$, one can execute $T$ steps for the subordinate levels ($s = 1, \dots, T$), producing sequences 
$$
(m_j(s))_{s=1}^T, \quad (r_j(s))_{s=1}^T
$$
for all $j \in I_i^-$.  
At each internal step, the actions from superior agents $a_j^+(1)$ are reused, keeping the top-level instructions constant throughout the execution of the internal sequence of steps.  

The resulting sequences of rewards and observations are then aggregated over time to form the inputs for $w_i$, for example:
$$
m_j(t) = \frac{1}{T} \sum_{s=1}^T m_j(s), 
\qquad 
r_j(t) = \sum_{s=1}^T r_j(s) .
$$

\subsection{Learning process}

We propose using reinforcement learning to train both agents’ policies as well as their communication and proxy-reward functions. Accordingly, the following Markov Decision Processes (MDPs) are defined for each agent. In what follows, we take agent $w_i$ as a reference. 

Each component of the agent — policy, communication, and proxy-reward function — can be formalized as a separate MDP. This formulation allows us to treat learning each component consistently within the reinforcement learning framework, while capturing the interactions among them.

\paragraph{Policy MDP.}  
The MDP for the agent's policy is defined as $\langle O_i, A_i, p_{\pi_i}, p_{\pi_i}^R, p_{\pi_i}^0 \rangle$, where:
\begin{itemize}
    \item $O_i = O_i^- \times O_i^+ = \prod_{j \in I_i^-} M_j \times \prod_{j \in I_i^+} A_j$ is the state space.
    \item $A_i$ is the action space.
    \item $p_{\pi_i}: O_i \times A_i \to \Delta(O_i)$ is an unknown transition function.
    \item $p_{\pi_i}^R: O_i \times A_i \to \Delta(\mathbb{R})$ is the reward distribution, determined by $R_i$, subordinate agents, and the external environment.
    \item $p_{\pi_i}^0 \in \Delta(O_i)$ is the initial state distribution, derived from the external environment and the communication functions of subordinate agents.
\end{itemize}

A trajectory for learning the policy of agent $w_i$ is
$$
\big(o_i(0), a_i(0), r_i^{\text{target}}(0), o_i(1), \dots, o_i(T)\big).
$$

\paragraph{Communication MDP.}  
Let $D_i = O_i^- \times \mathbb{R}^{l_i^-}$. The communication MDP is $\langle D_i, M_i, p_{\phi_i}, p_{\phi_i}^R, p_{\phi_i}^0 \rangle$, with:
\begin{itemize}
    \item state space: $D_i$
    \item action space: $M_i$
    \item transition function: $p_{\phi_i}: D_i \times M_i \to \Delta(D_i)$.
    \item reward distribution: $p_{\phi_i}^R: D_i \times M_i \to \Delta(\mathbb{R})$ as the reward distribution, determined by $R_i$, proxy-reward functions, and the external environment.
    \item $p_{\phi_i}^0 \in \Delta(D_i)$ as the initial state distribution, derived from the external environment and subordinate agents’ communication functions.
\end{itemize}

A trajectory for learning the communication function of agent $w_i$ is
$$
\big(d_i(0), m_i(0), r_i^{\text{target}}(0), d_i(1), \dots, d_i(T)\big),
$$
where $d_i(t) = (o_i^-(t), r_i^-(t-1))$.

\paragraph{Proxy-Reward Function.}  
Finally, for learning the proxy-reward function, the trajectory is
$$
\big(d_i(0), r_i(0), r_i^{\text{target}}(0), d_i(1), \dots, d_i(T)\big),
$$
with $\mathbb{R}$ as the action space.

This MDP-based formulation enables the usage of various RL and MARL algorithms for learning each component while accounting for interactions between the agent's policy, communication, and proxy-reward function.



\section{Application}

In the previous section, we formulated the Reinforcement Networks framework and described the inference procedure of the model. We now present one possible approach — among the wide variety of potential implementations — for realizing this framework. Our method builds upon the LevelEnv framework introduced in \citep{paolo2025tag}.

\subsection{Layered digraph}

The notion of \emph{outgoing depth} of a vertex in the graph $G$ is defined by induction.  First, for each sink vertex $w_s$, we set $d_s \coloneq 0$.
Next, consider a vertex $w_i$. Assume that the outgoing depths $d_j$ have already been defined for all $w_j \in V_i^-$.  
We then define $d_i \coloneq \max_{j \in I_i^-} d_j + 1$. In other words, the outgoing depth $d_i$ corresponds to the maximum length of a directed path from $w_i$ to any sink of the acyclic graph $G$.

We now introduce \emph{identity node}. Such nodes effectively act as transparent intermediaries within the graph, simply forwarding information and actions without modification. Formally, an identity node $w_n$ is a vertex satisfying the following conditions:
\begin{itemize}
    \item $l_n^- = 1$;
    \item $\phi_n(o_n^-, r_n^-) = (o_n^-, r_n^-)$, i.e., the communication and reward-proxy functions simply forwards the observations and reward from the subordinate agent;
    \item $\pi_n(o_n^-, a_n^+) = a_n^+$, i.e., the policy is deterministic and directly replicates the actions received from superior agents;
    \item $M_n = O_n^-$ and $A_n = A_n^+$, so that the message and action spaces coincide with the observation and instruction spaces, respectively.
\end{itemize}

We define the \emph{edge expansion} (by one node) operation for an edge $(w_i, w_j) \in E$ within a graph $G = (V, E)$, where $G \in \mathcal{G}$ and $\mathcal{G}$ denotes the set of DAGs. Formally, let $N = |V|$ and define a new identity node $w_{N+1}$ such that $I_{N+1}^+ = \{ i \}, \quad I_{N+1}^- = \{ j \}$.
The updated vertex set is $V' = V \cup \{w_{N+1}\}$, and the edge set is modified as $E' = \big(E \setminus \{(w_i, w_j)\}\big) \cup \{(w_i, w_{N+1}), (w_{N+1}, w_j)\}$. The resulting graph is defined as $G' = (V', E') \in \mathcal{G}$.

The expansion of an edge $(w_i, w_j)$ by $k \geqslant 2$ nodes is defined recursively as a sequence of single-node expansions. Specifically, first we expand the edge $(w_i, w_j)$ with identity node $w_{N+1}$, then at step $t = 1, \dots, k-1$, each edge $(w_{N+t}, w_j)$ is expanded by inserting an identity node.  

\begin{figure}
    \centering
    \includegraphics[width=0.8\textwidth]{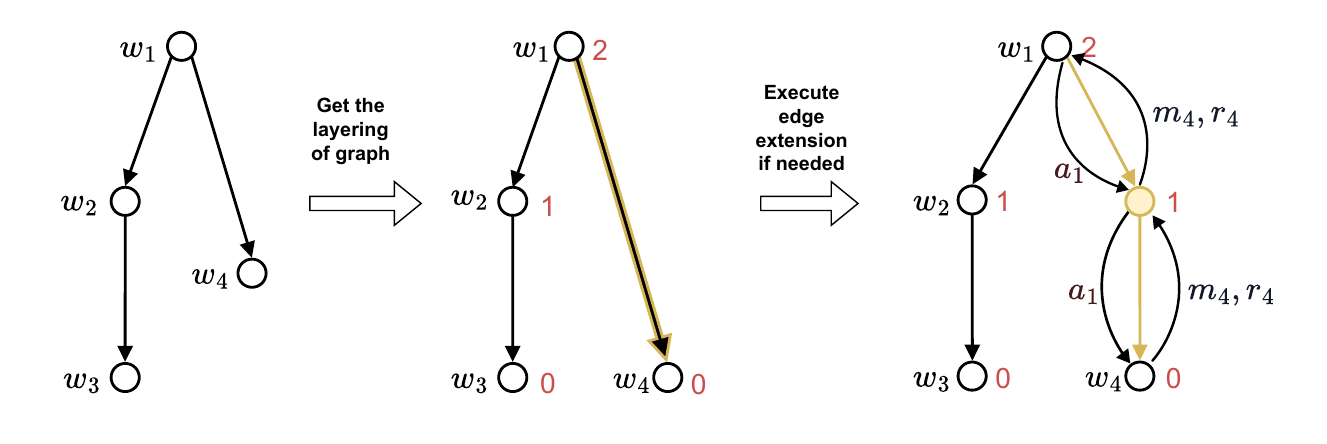}
    \caption{Transformation of a directed acyclic graph (DAG) into a \emph{layered digraph}.}
    \label{fig:layering}
\end{figure}

Namely, the operation inserts a new vertex $w_{N+1}$ along the edge $(w_i, w_j)$ that merely forwards received messages and actions without modification. See Figure ~\ref{fig:layering} for an illustration.

A \emph{layered digraph} is a directed acyclic graph (DAG) in which vertices are assigned to discrete layers such that every edge connects a vertex on one layer to a vertex on the immediately lower layer.  
Any directed acyclic graph (DAG) can be transformed into a functionally equivalent \emph{layered digraph}.  
To achieve this, it suffices to apply the edge expansion operation to every edge $(w_i, w_j) \in E$ for which $d_i - d_j > 1$, inserting $(d_i - d_j - 1)$ intermediate vertices.

The partitioning of a directed acyclic graph (DAG) into layers can be accomplished using algorithms such as the Longest Path algorithm, the Coffman-Graham algorithm, or the ILP algorithm of Gansner et al. (for an overview, see \citep{healy2001layer}).  
In this context, the edge expansion operation serves as an instrument to convert any DAG into a layered form, ensuring that each edge connects vertices on consecutive layers.  
The aforementioned algorithms can then be applied to determine an optimal placement of vertices across the layers, minimizing the total width or other layout-related objectives. However, all the graph sinks are restricted to be placed on the same level for the following application.

\subsection{Connection to \emph{LevelEnv}}
Transforming the DAG into a layered digraph enables a direct implementation of the \emph{LevelEnv} abstraction \citep{paolo2025tag}, where each layer acts as an environment for the superior level while functioning as a set of agents with respect to the subordinate one.

At layer $L_l$, each agent selects its action based on observations and rewards received from subordinate agents in $V_i^-$ together with directives provided by superior agents in $V_i^+$. The individual actions are then assembled into a joint action vector and passed to $L_{l-1}$, where masking ensures that each subordinate agent processes only the relevant components. Using its local observations and rewards, each subordinate agent produces outputs via the communication function $\phi$ (messages) and the function $\psi$ (rewards). These outputs are aggregated into layer-wide vectors and returned upward. At $L_l$, masking redistributes the returned signals to the appropriate agents, after which $\phi$ and $\psi$ are applied again to produce the aggregated outputs for the superior layer.

Thus, layers exchange information exclusively through unified vectors, while masking enforces the graph structure by restricting each agent to signals from its designated neighbors. This design simultaneously preserves agent-level independence and allows each layer to operate on its own temporal and spatial scale.

\section{Experiments}
To assess the benefits of Reinforcement Networks relative to TAG-based systems with policies trained at each vertex, we perform experiments on two MARL tasks: \textit{MPE Simple Spread} and  a discrete version of \textit{VMAS Balance}. All experiments use four agents. We compare our method with two baselines: IPPO \cite{dewitt2020ippo} and 3PPO \cite{paolo2025tag}. IPPO corresponds to a four-node DAG with no edges, in which each agent is trained independently via PPO; this architecture is expressible within the Reinforcement Network framework. 3PPO comprises a three-level hierarchy: four motor nodes, two mid-level controllers (each supervising two motors), and a single top-level controller. This structure forms a full binary tree of height three and can also be encoded as a Reinforcement Network.

Our proposed \textbf{bridged-3PPO} architecture augments 3PPO with skip-connections linking the top and bottom-level nodes. Representing these skip-connections in TAG requires introducing identity vertices. For all multi-level configurations, higher layers act only every two timesteps. The reward-proxy function is instantiated as an averaging operator, and communication is implemented via the identity map. We further evaluate \textbf{bridged-3PPO-comm}, a variant in which the communication function $\phi$ is learned using the autoencoder training method from \citep{paolo2025tag}.

\begin{figure}[h]
\centering
\subcaptionbox{}{\includegraphics[width=0.45\textwidth]{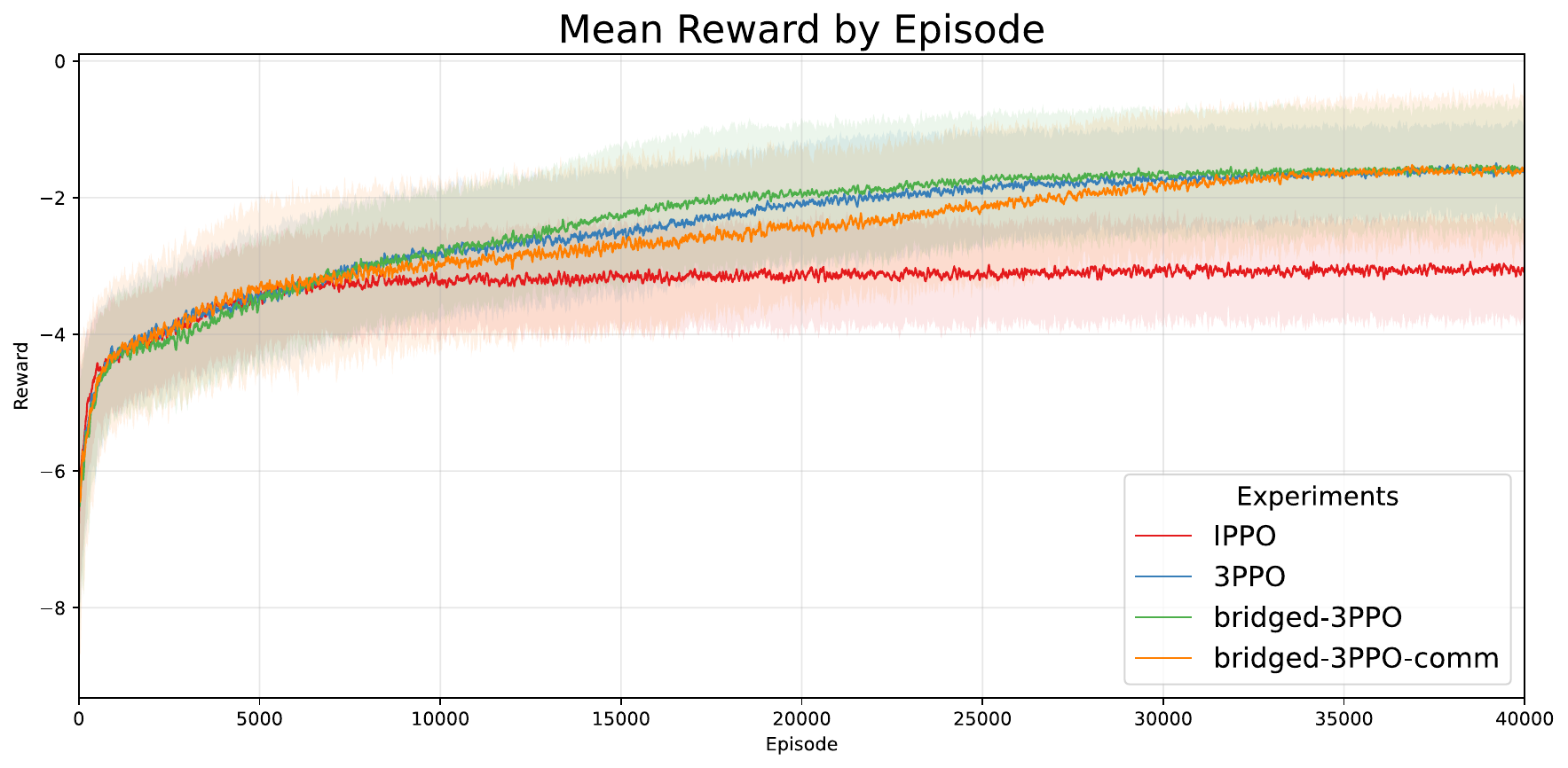}}%
\hfill 
\subcaptionbox{}{\includegraphics[width=0.45\textwidth]{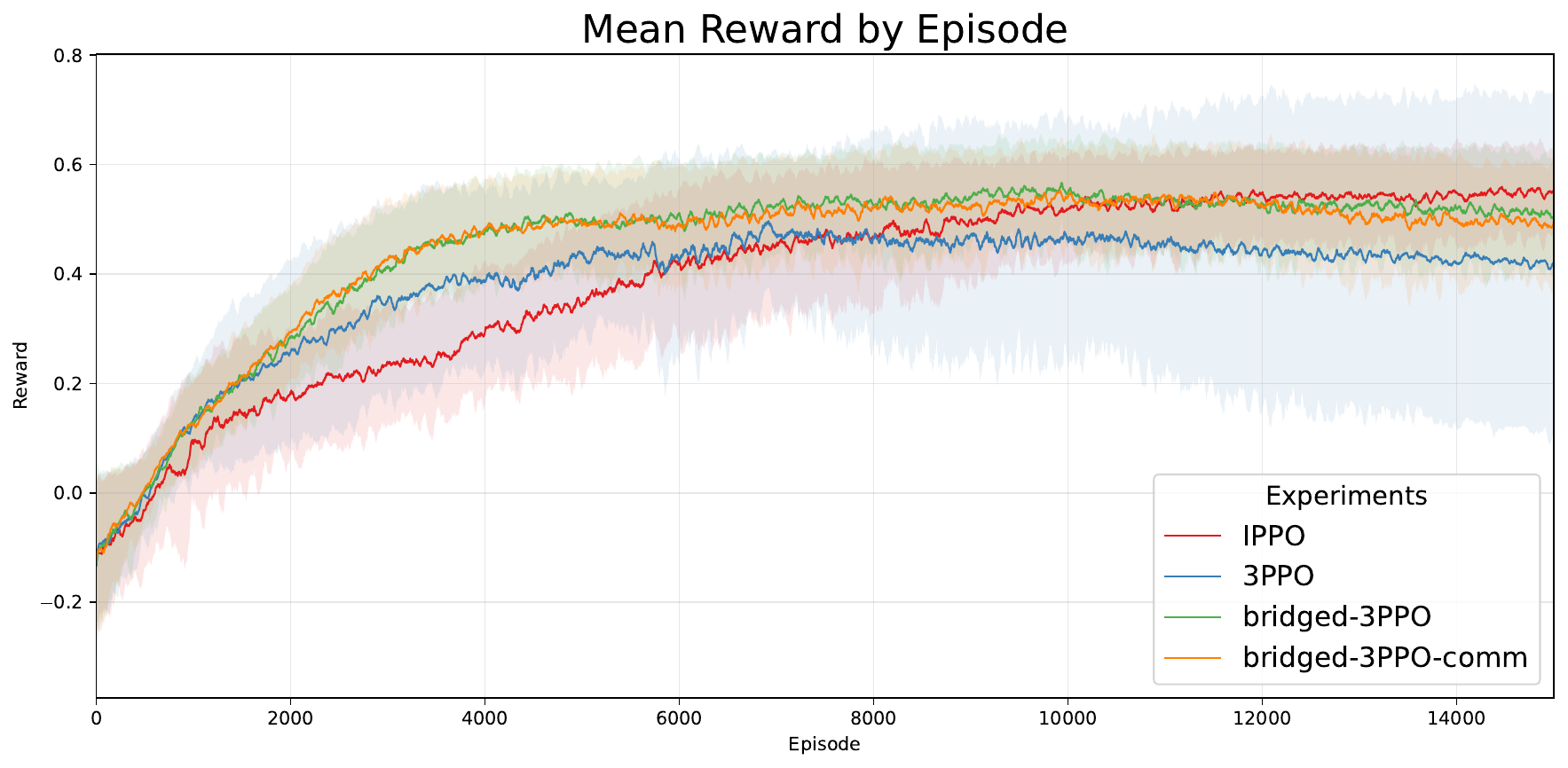}}%
\caption{Mean episode reward in the MPE Simple Spread (a) and VMAS Balance (b) environments. Results are averaged over 5 random seeds. Shaded regions denote 95\% confidence intervals.}
\label{fig:simple_spread}
\end{figure}

Training curves are reported in Figure~\ref{fig:simple_spread}.  
Results are averaged over five random seeds.

\subsection{MPE Simple Spread}

In the \textit{MPE Simple Spread} environment, agents must coordinate in a 2D plane to cover all landmarks while avoiding collisions. This task requires cooperative spatial reasoning and benefits strongly from structured information flow.

IPPO serves as a non-hierarchical baseline: although each agent observes the relative positions of others, the absence of a coordinating structure yields unstable and low-reward behavior. All hierarchical variants substantially outperform IPPO, demonstrating the importance of multi-level reasoning even with simple directive and communication mechanisms.

Across hierarchical methods, \textbf{bridged-3PPO} exhibits a brief warm-up dip and slightly increased variance relative to 3PPO. This effect is expected: adding bridges increases the state and directive dimensionality at higher levels, temporarily complicating credit assignment early in training. However, \textbf{bridged-3PPO consistently converges faster}, reaching its performance plateau in fewer episodes than 3PPO. This suggests that skip-connections provide useful shortcuts for information flow, allowing high-level agents to react more quickly to bottom-level dynamics.

Introducing learned communication in \textbf{bridged-3PPO-comm} further reduces the warm-up phase by compressing observations into a learned latent representation. While the additional learning burden slightly slows the final convergence rate, the improved early-training stability indicates that learned feature extraction can mitigate part of the state explosion introduced by bridging.

\subsection{VMAS Balance}

In the \textit{VMAS Balance} task, a team of agents must cooperatively stabilize and lift a spherical package supported by a line under gravity, delivering it to a goal region. All agents receive an identical reward based on changes in package - goal distance, and incur a large penalty if the package or the line touches the floor. The task is highly coordination-sensitive, as even a single misaligned action can destabilize the system.

Surprisingly, IPPO reaches higher final returns than 3PPO, indicating that excessive hierarchical abstraction may restrict the fine-grained corrective behavior needed for this domain. However, \textbf{bridged-3PPO} closes the performance gap and reaches the same reward plateau using only about two-thirds of the interaction steps required by IPPO, highlighting a sample-efficiency benefit of bridging even when pure hierarchical decomposition underperforms.

The communication-enabled variant \textbf{bridged-3PPO-comm} further compresses each agent's observation space, which reduces the training burden on higher-level agents without degrading final performance.

The \textbf{most robust effect across both variants is a marked reduction in reward variance} (see Fig.~\ref{fig:simple_spread}). While 3PPO shows oscillatory learning, both bridged architectures exhibit substantially smoother reward trajectories. \footnote{Reward curves are reproduced from the training logs.} Lower variance indicates that links across non-adjacent layers help stabilize training.

Overall, even when bridges do not improve peak performance, they consistently improve \textit{stability} and \textit{sample efficiency}, two properties critical for scaling cooperative multi-agent systems. These findings support our hypothesis that structured skip-connections across hierarchical levels are a helpful architectural feature for hierarchical MARL.

\subsection{Computational Resources}

The research was carried out using the MSU-270 supercomputer of Lomonosov Moscow State University.

\section{Conclusion}
In this work, we propose a novel, flexible, and scalable approach to constructing solutions for collaborative MARL tasks. Our main contribution is a unified framework for hierarchical MARL that generalizes a wide range of existing models and substantially simplifies the use of DAG-structured hierarchies compared to the TAG formalism. Reinforcement Networks introduce additional degrees of freedom in inter-agent communication while preserving well-defined intra-agent MDPs. We argue that connections across vertices at different depths can improve both training stability,  and sample efficiency. This framework also opens multiple avenues for advancing the theory and practical design of Reinforcement Networks.

Below, we outline several promising avenues for future work.

\paragraph{Optimal topology construction.} Developing algorithms and methods to identify optimal DAG topologies tailored to specific MARL tasks is a key challenge. Progress in this area would support both researchers and practitioners in designing more effective systems.

\paragraph{Proxy-reward and communication function training.} Advancing the theory and practice of training proxy-reward and communication functions is, in our view, one of the most promising directions for enhancing both the robustness and performance of our approach. 

\paragraph{LLM-based agents.} Another important direction is to investigate how our methods can be applied to the tuning of LLM-based agents. This raises a number of open questions, ranging from the design of communication and proxy-reward functions to optimizing training and inference efficiency.

\section*{Acknowledgments}

This work was supported by the Ministry of Economic Development of the Russian Federation
in accordance with the subsidy agreement (agreement identifier 000000C313925P4H0002;
grant No 139-15-2025-012).

\bibliography{references}

\appendix

\section{Graph layering}

Pseudocode of the proposed DAG conversion to layered digraph algorithm might be found at ~\ref{alg:layering}. 

\begin{algorithm}[h]
\caption{Graph Transformation with Outgoing Depth Computation}
\begin{algorithmic}
\State \textbf{Input:} Directed graph $G = (V, E)$
\State \textbf{Output:} Layered digraph $G' = (V', E')$ with identity nodes

-----------------------------
\Procedure{OutgoingDepth}{$G$}
  \ForAll{$w_i \in V$}
    \State $d_i \gets$ \textbf{None}
  \EndFor
  \ForAll{$w_i \in V$}
    \If{$d_i =$ \textbf{None}}
      \State \Call{DFS-Visit}{$w_i$}
    \EndIf
  \EndFor
\EndProcedure

\Procedure{DFS-Visit}{$w_i$}
  \State $d_i \gets 0$
  \State ChildrenDepths $\gets$ [ ]
  \ForAll{$w_j \in I_i^-$} \Comment{$I_i^-$ = set of children of $w_i$}
    \If{$d_j =$ \textbf{None}}
      \State \Call{DFS-Visit}{$w_j$}
    \EndIf
    \State ChildrenDepths.append($d_j$)
  \EndFor
  \If{ChildrenDepths $\neq$ [ ]}
    \State $d_i \gets \max(\text{ChildrenDepths}) + 1$
  \EndIf
\EndProcedure

\Procedure{EdgeExtension}{$G, w_i, w_j$}
  \State $N \gets |V|$
  \State introduce new node $w_{N+1}$
  \State $A_{N+1} \gets A_j$ \Comment{copy attribute of $w_j$}
  \State $M_{N+1} \gets M_i$ \Comment{copy attribute of $w_i$}
  \State $E' \gets (E \setminus \{(w_i, w_j)\}) \cup \{(w_i, w_{N+1}), (w_{N+1}, w_j)\}$
  \State $V' \gets V \cup \{w_{N+1}\}$
  \State \Return $G' = (V', E')$
\EndProcedure

\Procedure{TransformGraph}{$G$}
  \State \Call{OutgoingDepth}{$G$} \Comment{compute outgoing depths $d_i$ for all $w_i \in V$ with the algorithm you like}
  \State $V' \gets V$, $E' \gets E$
  \ForAll{$(w_i, w_j) \in E$}
    \If{$d_i - d_j > 1$}
      \State $k \gets 1$
      \State $(V', E') \gets$ \Call{EdgeExtension}{$(V', E'), w_i, w_j$}
      \While{$k < d_i - d_j - 1$}
        \State $(V', E') \gets$ \Call{EdgeExtension}{$(V', E'), w_{|V'|}, w_j$}
        \State $k \gets k + 1$
      \EndWhile
    \EndIf
  \EndFor
  \State \Return $(V', E')$
\EndProcedure

\end{algorithmic}

\label{alg:layering}
\end{algorithm}

\section{Experimental Setup}

We use a modified \verb|pytorch| implementation of TAG \cite{paolo2025tag}. Environments from \verb|PettingZoo| and \verb|VMAS| packages are used.

For three-level hierarchies, we use the following action frequencies: middle and top levels act every two timesteps of the bottom level.

\section{Hyperparameters}

\subsection{Actor--Critic Network Architecture}

All PPO-based agents use identical actor and critic architectures. The hyperparameters are summarized below:

\paragraph{Network structure.}
\begin{itemize}
    \item \textbf{Number of layers:} 3 (actor), 3 (critic)
    \item \textbf{Input layer:} Observation size $\rightarrow 64$
    \item \textbf{Hidden layers:} $64 \rightarrow 64$ with Tanh activation
    \item \textbf{Output layer:} 
          \begin{itemize}
              \item Actor: $64 \rightarrow \text{Action size}$
              \item Critic: $64 \rightarrow 1$
          \end{itemize}
    \item \textbf{Activation functions:} Tanh for all intermediate layers
    \item \textbf{Output initialization:} 
          \begin{itemize}
              \item Actor std: $0.01$
              \item Critic std: $1.0$
          \end{itemize}
    \item \textbf{Action type:} Discrete
\end{itemize}

\subsection{Training Hyperparameters for 3PPO}

\begin{itemize}
    \item \textbf{Total training steps:} 4,000,000
    \item \textbf{Learning rate:} 0.001
    \item \textbf{Learning rate annealing:} Enabled
    \item \textbf{Max gradient norm:} 0.5
    \item \textbf{Buffer size:} 2,048
    \item \textbf{Number of minibatches:} 8
    \item \textbf{Update epochs:} 4
    \item \textbf{Discount factor $\gamma$:} 0.99
    \item \textbf{GAE $\lambda$:} 0.95
    \item \textbf{Advantage normalization:} Enabled
    \item \textbf{Clip coefficient:} 0.1
    \item \textbf{Value loss clipping:} Enabled
    \item \textbf{Entropy coefficient:} 0.01
    \item \textbf{Value function coefficient:} 0.5
    \item \textbf{Target KL:} 0.015
\end{itemize}

\subsection{Training Hyperparameters for IPPO}

\begin{itemize}
    \item \textbf{Total training steps:} 4,000,000
    \item \textbf{Learning rate:} $2.5 \times 10^{-4}$
    \item \textbf{Learning rate annealing:} Enabled
    \item \textbf{Discount factor $\gamma$:} 0.99
    \item \textbf{GAE $\lambda$:} 0.95
    \item \textbf{Batch size:} 2,048
    \item \textbf{Number of minibatches:} 4
    \item \textbf{Update epochs:} 4
    \item \textbf{Advantage normalization:} Enabled
    \item \textbf{Clip coefficient:} 0.2
    \item \textbf{Value loss clipping:} Enabled
    \item \textbf{Entropy coefficient:} 0.0
    \item \textbf{Value function coefficient:} 0.5
    \item \textbf{Max gradient norm:} 0.5
    \item \textbf{Target KL:} None
\end{itemize}

\subsection{Autoencoder Hyperparameters}

The autoencoder used for communication learning is defined as follows:

\paragraph{Encoder.}
\begin{itemize}
    \item Input: Observation shape $\rightarrow 32$
    \item Activation: ReLU
    \item Output: $32 \rightarrow 8$
\end{itemize}

\paragraph{Decoder.}
\begin{itemize}
    \item Input: $8 \rightarrow 32$
    \item Activation: ReLU
    \item Output: $32 \rightarrow \text{Observation shape}$
    \item Final activation: None
\end{itemize}

\paragraph{Training.}
\begin{itemize}
    \item Loss: Mean squared error (MSE)
    \item Number of epochs: 50
\end{itemize}

For all hierarchical configurations (3PPO, bridged-3PPO, and bridged-3PPO-comm): each agent outputs a discrete directive, with a distinct directive produced for each subordinate agent.

For the \textbf{bridged-3PPO-comm} configuration specifically:
\begin{itemize}
    \item The communication embedding space is fixed to dimension 12 at all hierarchy levels.
    \item The communication function is implemented using the autoencoder described above.
\end{itemize}


\end{document}